\documentstyle[epsf,prl,aps,twocolumn]{revtex}
\input epsf
\begin{document}
\draft
\title
{Doped planar quantum antiferromagnets
with striped phases}
\author{A.H. Castro Neto$^{1,2}$ and Daniel Hone$^{1}$}
\bigskip
\address{$^{1}$
Institute for Theoretical Physics,
University of California,
Santa Barbara, California, 93106}

\address{$^{2}$
Department of Physics,
University of California,
Riverside, California, 92521 }
\date{\today}

\maketitle

\bigskip

\begin{abstract}
We study the properties of the striped phases that have been proposed
for the doped cuprate planar quantum antiferromagnets.
We invoke an effective, spatially
anisotropic, non-linear sigma model in two space dimensions.  Our theoretical
predictions are in {\it quantitative} agreement with recent experiments.
We focus on (i) the staggered magnetization at $T=0$ and (ii) the N\'eel
temperature, as functions of doping; these have
been measured recently in La${}_{2-x}$ Sr${}_x$ Cu O${}_4$ with
$0 \leq x \leq 0.018$.
Good agreement
with experiment is obtained using parameters determined previously and
independently for this system.
These results support the proposal
that the low doping (antiferromagnetic) phase of the cuprates has a
striped configuration.
\end{abstract}


\pacs{75.10Jm, 74.72.Dn, 75.50Ee, 64.75.+g}

\narrowtext

Our understanding of the {\it undoped} planar cuprates related to high
temperature superconductors has been informed largely by
the insights of Chakravarty, Halperin and Nelson \cite{chn}
(hereafter, CHN). It is now
widely accepted that
the antiferromagnetic phase of these materials can be well described by an
effective non-linear sigma model.
However, the situation is not so clear when we turn to the
behavior of doped systems, as the antiferromagnetic moves toward a
superconducting instability.   It has long been recognized
\cite{outros,review} that there is a
tendency for the holes to phase separate in these strongly correlated
insulators. Experimental observations \cite{experimental} support this
scenario for some of the La-cuprates at high concentrations of dopants.
In addition,
recent nuclear quadrupole resonance (NQR) and muon spin resonance ($\mu$SR)
experiments\cite{borsa} in La$_{2-x}$Sr$_x$CuO$_4$, with
$0\leq x \leq 0.018$, have been interpreted within a picture where holes are
segregated into a set of parallel stripes.  It is the primary goal of this
Letter, in fact, to do a more careful
theoretical analysis of the consequences of that model and to compare the
predictions to those experiments.

With the xy coordinate axes chosen as the crystal axes in the CuO$_2$ plane,
we take an
array of equivalent uniformly spaced stripes
parallel to the y-axis.  In this simplest of models
we neglect both static and dynamic fluctuations (and domain formation).
The situation
is not very different if we consider diagonal stripes such as the ones found
in La$_2$NiO$_{4.125}$ and La$_{1.8}$Sr$_{0.2}$NiO$_4$ \cite{nickel}.
But here we will treat only the horizontal stripe geometry suggested for
La$_{2-x}$Sr$_x$CuO$_4$ by the experimental data \cite{stripe}.
In the regions between the stripes
the Cu spin correlations are presumed to remain
antiferromagnetic.

 As in the CHN analysis~\cite{chn} of the pure case, we will argue that
the underlying symmetries suggest an appropriate nonlinear sigma model to
describe the long wavelength behavior which determines the phase diagram.
We start from a microscopic model with
the spins interacting via Heisenberg antiferromagnetic exchange in the regions
 between stripes, and by a weaker such exchange
across each stripe.  Conceptually, we use real space renormalization
in the x-direction to integrate out the short distance dynamics,
starting with blocks the size of the region between stripes.
This is continued to a scale of the correlation length, much larger
throughout the region of interest to us than the inter-stripe spacing.
The expected leading result for the low frequency dynamics in which we
are interested is a spatially anisotropic nearest neighbor
Heisenberg exchange model,
\begin{eqnarray}
H&=& J_y \sum_{n,<m,m'>} \vec{S}(n,m) \cdot \vec{S}(n,m')
\nonumber
\\
&+& J_x \sum_{m,<n,n'>} \vec{S}(n,m) \cdot \vec{S}(n',m)
\label{heisenberg2}
\end{eqnarray}
where $n$ and $m$ label the sites in the x and y directions, respectively.
$J_x$ and $J_y$ are the effective spin exchange strengths in the two
directions.
In general they depend on the doping $x$, but the exact form of that
dependence relies, of course, on microscopic details.

The destruction of
long range order at zero temperature and the suppression of the
staggered magnetization with increased hole doping then come about from
the dimensional crossover from two to one dimensional behavior
with
the increasing anisotropy of the system (in our effective $\sigma$-model
the critical coupling constant $g_c$, which separates ordered from disordered
behavior at $T=0$, decreases with doping).
At the same time, the renormalized spin stiffness decreases as the
density of holes grows.
As $g_c(x)$ decreases it equals at some critical
concentration $x_c$ the coupling constant of the undoped system
and antiferromagnetism disappears.

To simplify the study of the long wavelength physics it is convenient to
take the continuum limit of
(\ref{heisenberg2}).
We use the usual spin coherent state representation \cite{spincoh} and
we find
that the effective action (in euclidean time) in the partition function
can be written as
\begin{eqnarray}
S_{eff} &=& \frac{1}{2}
\int_0^{\beta \hbar} d \tau \int dx \int dy
\left\{ S^2 \left[J_y \left(\partial_y \hat{n}\right)^2
+ J_x \left(\partial_x \hat{n}\right)^2 \right] \right.
\nonumber
\\
&+& \left.
\frac{\hbar^2}{2 a^2 (J_x+J_y)} \left(\partial_{\tau} \hat{n}\right)^2
\right\},
\label{effa1}
\end{eqnarray}
where $\hat n$ is a unit vector and $a$ is the sub-lattice constant.
This gives spin wave velocities
$c_y^2 = 2 S^2 a^2 J_y (J_x+J_y)/\hbar^2$ and
$c_x^2 = 2 S^2 a^2 J_x (J_x+J_y)/\hbar^2$,
which agree with a simple (noninteracting) spin wave calculation based on
(\ref{heisenberg2}). This action was studied numerically
in a different context a few years ago \cite{parola}.

It is useful to rewrite (\ref{effa1}) more symmetrically by a dimensionless
rescaling of the variables:
$x'= (J_y/J_x)^{1/4} x \Lambda$,
$y'= (J_x/J_y)^{1/4} y \Lambda$ ($\Lambda$ is a momentum cut-off), and
$\tau'= \sqrt{2 (J_x+J_y) \sqrt{J_x J_y}} S a  \tau/\hbar$.
Then the effective action (\ref{effa1}) becomes
\begin{equation}
S_{eff} = {\hbar\over (2 g_0)} \int_0^{\hbar \Lambda \beta c_0} d \tau' \int
dx' \int dy'
\left(\partial_{\mu} \hat{n}\right)^2,
\end{equation}
where $\mu$ takes the values $ x',y',\tau'$,
$g_0 = \hbar c_0 \Lambda/\rho^0_s =
\left[2 (J_x+J_y)/\sqrt{J_x J_y}\right]^{1/2}(a \Lambda)/S$
is~the~bare~coupling~constant, $c_0= [2 (J_x+J_y) \sqrt{J_x J_y}]^{1/2}
(a S)/\hbar$
the spin wave velocity and
$\rho^0_s = \sqrt{J_x J_y} S^2$ the classical spin stiffness of the rescaled
model. The original anisotropy is now hidden in the limits.
We started with a problem with a
finite bandwidth, a lower bound on length which requires us to
impose a cutoff in
the original continuum formulation.  The change of variables introduces
an anisotropy in the cutoffs.

The $\sigma$-model action, and the spin correlations it implies, can be
studied in the large $N$ limit ($N$ is the number of components of
$\hat n$), where a saddle point approximation becomes exact \cite{largeN,assa}.
The staggered spin-spin correlation function is given by
$\Xi(\vec{k},\omega_n) = g_0/(k^2 + \omega_n^2 +m^2)$
where $m$ is defined by the self-consistent condition,
$\sum_n \sum_{\vec{k}} \Xi(\vec{k},\omega_n;m) = 1$,
and sets the scale for the correlations in the system.
The self-consistency equation can also be written, after the
sum over Matsubara frequencies, as
\begin{equation}
\frac{g_0}{2} \int \frac{d^2 k}{(2 \pi)^2} \frac{\coth\left(
\sqrt{k^2+m^2} \beta c_0 \Lambda/2 \right)}{\sqrt{k^2+m^2}} = 1
\label{mats}
\end{equation}
Formally, this integral has a logarithmic ultraviolet divergence.
If we choose an isotropic cutoff $~ \pi/\Lambda a$
for the Fourier transform of the original problem (\ref{effa1}),
then $|k_x| < \pi \alpha^{1/4}/(\Lambda a)$ and
$|k_y| < \pi \alpha^{-1/4}/(\Lambda a)$, where
\begin{equation}
\alpha = J_x / J_y,
\label{anisotropy}
\end{equation}
is the anisotropy parameter (as usual,
we take $\Lambda a = 2 \sqrt{\pi}$ to preserve the area of the magnetic
Brillouin zone).
By the definition above, the coupling constant
$g_0(\alpha) = g_0(1) \sqrt{(1+\alpha)/(2 \sqrt{\alpha})}$.
In what follows we will choose $g_0(1)$ renormalized to give the
{\it interacting} spin wave velocity of Oguchi~\cite{oguchi},
rather than the simple spin wave value above.
Because the effective exchange $J_x$ in the x-direction is weakened by the
stripes, we have $0\leq \alpha \leq 1$.

At high anisotropy ($\alpha \to 0$) the spin chains
become disconnected and we find one dimensional behavior.
In this limit  $m >> \alpha^{1/4}$, and we find at zero temperature
 from (\ref{mats}),
$m(\alpha) \approx 4  \alpha^{-1/4}/\sqrt{\pi}
\exp\left[-4 \pi^{3/2}/(\alpha^{1/4}
g_0(\alpha))\right],$
which is just the Haldane gap \cite{haldane}.
We came to this result because the
non-linear $\sigma$-model without a topological term describes the
behavior of an {\it integer} spin chain \cite{gap}.
In the opposite limit, $m << \alpha^{1/4}$,
the system is fully two dimensional
and Eq. (\ref{mats}) at zero temperature leads to
$m(\alpha) \approx 4  \sqrt{2} \pi \alpha^{1/4}/\sqrt{1+\alpha}
\left[1/g_c(\alpha)-1/g_0(1)\right]$
where
\begin{eqnarray}
g_c(\alpha) &=&
 8 \pi^{3/2} \sqrt{\alpha/(1+\alpha)}  \left\{
 \ln\left(\sqrt{\alpha}+\sqrt{1+\alpha}\right)\right.
\nonumber
\\
&+& \left. \sqrt{\alpha}\ln[(1+\sqrt{1+\alpha})/\sqrt{\alpha}]
\right\}^{-1}
\label{gca}
\end{eqnarray}
is the critical coupling constant of the theory. It is easy to see that
$g_c$ decreases monotonically to zero
as the system becomes more anisotropic and
the increasing quantum fluctuations again make magnetic order less stable.
Dimensional crossover occurs when the correlation length in the x direction
becomes of the order of the lattice spacing, that is,
$m(\alpha_t) \approx \alpha_t^{1/4}$, and a
numerical estimate gives $\alpha_t \approx 0.001$.

To understand the corrections to the classical limit we employ
a renormalization group (RG) calculation up to second order in the coupling
constant.  Proceeding as in \cite{chn}, we find within a one-loop
approximation the renormalized spin stiffness
\begin{eqnarray}
\rho_s(\alpha) \approx
\rho^0_s(\alpha) \left(1-g_0(1) / g_c(\alpha)\right),
\label{stiff}
\end{eqnarray}
reduced from its classical value $\rho_s^0$ for fixed $\alpha$.
As a first consequence we see that the stiffness will vanish
for some critical value, $\alpha = \alpha_c$, which
can be calculated numerically. Oguchi's spin wave theory \cite{chn,oguchi}
gives $g_0(1) \approx 9.536$.
Then (\ref{stiff}) gives $\alpha_c \approx 0.047$, much greater than
$\alpha_t$, so the transition from classical to quantum
regimes (ordered to disordered ground state) occurs well before
the potential crossover from two to one dimensional behavior \cite{comment}.

We turn to the finite temperature behavior.  Again, we can use the
inequality $m\ll\alpha^{1/4}$ in estimating  the correlation length
$\xi \propto 1/m$ from equation (\ref{mats}).
However, we know from studies of the
undoped system that the one loop calculation does not give a very good
result there for the prefactor of the temperature dependent exponential
expression for $\xi$, and the same will surely be true for $\alpha<1$.
Here we use an interpolation formula between the
exact result of Hasenfratz and Niedermayer \cite{exact}
for the nonlinear sigma model, which is valid close to the ordered phase,
and the result for the
renormalized critical regime where $\xi \propto T^{-1}$ \cite{chn}:
\begin{equation}
\xi(T,\alpha) \approx \frac{e}{8} \frac{\hbar c_0}{2 \pi \rho_s(\alpha)}
\frac{e^{2 \pi \rho_s(\alpha)/k_B T}}{
\left(1+\frac{k_B T}{4 \pi \rho_s(\alpha)}\right)}.
\label{interpol}
\end{equation}

In general the staggered magnetization depends on the short, as well as
on the long wavelength physics of the problem.  As in \cite{chn},
instead of using the non-linear sigma model directly we observe that
the equal time staggered spin-spin correlation function at large
distances becomes proportional to the square of the staggered magnetization,
$\lim_{x \to \infty,y \to \infty} \Xi(x,y) = \left(M_s/M_0\right)^2$,
where $M_0$ is the staggered magnetization for the classical problem
(the fully aligned N\'eel state, without quantum fluctuations).
At zero temperature, in the ordered phase, the correlation length is
infinite. The only scale
left is the Josephson length, $\xi_J = \hbar c_0/\rho_s$.
On one hand the
asymptotic (large distance) limit of the correlation function has been
established at this scale. On the other hand this is the maximum
scale at which the
long wavelength dominated theory we have developed can be trusted.
Therefore, we equate the correlation function at this (scaled) length
to the square of the relative zero point magnetization and
using the RG result (\ref{stiff}), we find
\begin{equation}
\frac{M_s(\alpha)}{M_s(1)} = \sqrt{\frac{1-g_0(1)/g_c(\alpha)}{
1-g_0(1)/g_c(1)}}.
\label{ms}
\end{equation}
As expected, $M_s(\alpha_c)=0$, and close to $\alpha_c$ we find
$M_s(\alpha) \approx \left(\alpha/\alpha_c-1\right)^{1/2}$
which has the mean field exponent of $1/2$.

To have true long range order at $T>0$ we need to invoke
the weak coupling between planes, $J_{\perp}$. Because
$J_{\perp}/J \approx 5\times 10^{-5}$ \cite{keimer} is so small,
long range correlations have built up in the plane well above the
ordering temperature $T_N$.  Spin fluctuations then involve large
correlated regions, and it is a good approximation to treat
$J_\perp$ within a mean field theory.
To determine the dependence of the N\'eel temperature on anisotropy one needs
to know the connection between the classical spin stiffness
and $\alpha$. Here we assume as a simple model that the classical stiffness is
approximately independent of $\alpha$ over the region of interest,
equal to the undoped value $\rho_s^0(\alpha=1)$, so
$\sqrt{J_x J_y} = J$.

Magnetic order is destroyed by thermal fluctuations, at a
temperature $T_N$, when the energy $k_BT_N$ becomes
sufficient to flip the spins in a region of linear dimension of the order of
the coherence length $\xi$.
Since the number of spins in this region is proportional to $(\xi/a)^2$,
and the relative staggered magnetization in the
region is given by $M_s/M_0$, we estimate
\begin{equation}
k_B T_N(\alpha) \approx
J_{\perp} \left(\frac{\xi(T_N,\alpha)}{a} \frac{M_s}{M_0}\right)^2.
\label{heur}
\end{equation}
This expression has been used previously~\cite{chn} to
estimate $J_\perp/k_B \approx 0.01$ K from the experimental
$T_N$ of the pure material.
Since $M_s/M_0 < 1$, this gives $(\xi/a) > 10$ for $T_N > 1$ K,
suggesting that this mean field theory is reasonable for $T_N$ greater than
a few kelvin.
We now substitute the expression
(\ref{interpol}) for $\xi$ in the quantum critical regime into (\ref{heur})
to find
$T_N(\alpha) \approx \left(\alpha/\alpha_c-1\right)^{1/3}$,
defining a critical exponent of $1/3$.

In order to compare our results with the experiments
we need the relationship between
the anisotropy parameter (\ref{anisotropy}) and the doping concentration $x$.
Explicit calculation of that is complicated, depending intrinsically
on the chemistry of the material; it goes beyond our previous
analysis. Instead, we propose a simple parametrization,
\begin{equation}
\alpha(x) = e^{-x/x_0},
\label{alpha}
\end{equation}
with $x_0$, the single (non-universal) parameter characterizing the behavior,
dependent in an unkown way on the microscopic details of
the system. We fix the value of $x_0$
from experiment.  We stress that this is the {\it only}
free parameter in our theory. All others are obtained independently for
the undoped material, and they are well known.

\begin{figure}
\epsfysize=3.8 truein
\epsfxsize=3.2 truein
\centerline{\epsffile{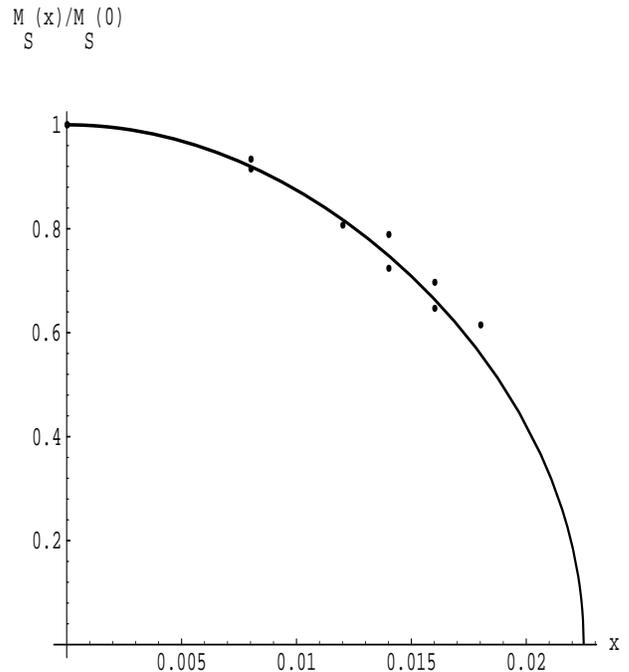}}
\caption[]{Staggered magnetization (normalized relative to the undoped case)
as a function of doping (line) and experiment (dots).}
\label{stagmag}
\end{figure}

Experimentally the staggered magnetization
and the N\'eel temperature vanish at a critical doping $x_c \approx
0.02 - 0.023$ \cite{borsa}.
This value of $x_c$ must correspond in our model to the point
at which the spin wave stiffness vanishes, that is,
to the critical value $\alpha_c \approx 0.047$.
Then (\ref{alpha}) gives $x_0 \approx  0.007$.
Depending on the density of holes in the stripes, this corresponds to a
stripe spacing of order 10 lattice constants or so,
a reasonable scale on which to expect substantial changes to occur.
Substituting (\ref{alpha}) into (\ref{ms}),
we find the staggered magnetization as a function of
doping. In Fig. (\ref{stagmag}) we show the theoretical prediction
(full line)
and the experimental values (dots) from ref. \cite{borsa}.
In view of the approximations used in the theory above,
the agreement is impressive.

\begin{figure}
\epsfxsize=3.2 truein
\epsfysize=3.8 truein
\centerline{\epsffile{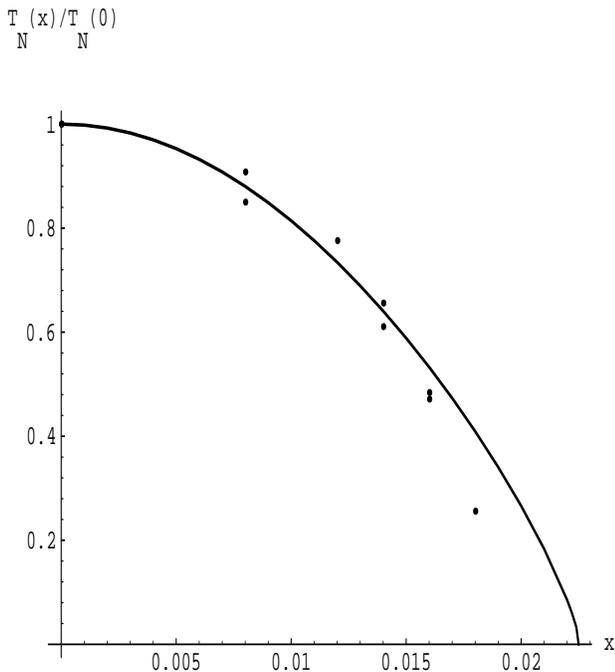}}
\caption[]{N\'eel temperature (normalized relative to the undoped case)
as a function of doping. Theory (continuous line) and experiment (dots).}
\label{tncurve}
\end{figure}

The calculation of the N\'eel temperature from (\ref{heur})
brings in new parameters, including the interplanar exchange $J_{\perp}$,
but they are all set by those of the undoped material.
We need no additional assumptions; we simply use the same parameters
taken~\cite{chn} by CHN to fit experiment:
$\rho_s(1)/k_B \approx 187 $ K and
$\hbar c_0(1) \approx 7.7 \times 10^3 $ A$^o$  K.
Using equations (\ref{interpol}) and (\ref{heur}) we
obtain the full curve in Fig. \ref{tncurve}. The agreement
between theory and experiment is very good except for the smallest value
of $T_N$, and can be improved
by small adjustments of the parameters \cite{next}.
Our conclusion is, therefore, that
our theory is fully consistent with the recent experiments on
La$_{2-x}$ Sr$_x$ Cu O$_4$.

In summary, we have developed an effective field theory for magnetic
correlations in doped antiferromagnets that are anisotropic due to the stripe
correlations expected from microscopic phase segregation of holes. We suggest
that the long wavelength spin fluctuations
are described by a spatially anisotropic non-linear sigma model.
The experimental parameter, doping concentration $x$,
is reflected in the theory by the magnetic exchange anisotropy $\alpha$,
taken to depend exponentially on $x$ (introducing the only free parameter of
the theory, the decay rate $x_0$).
We further assume that the classical spin stiffness is not affected
by doping.
We believe that the good
agreement between this theory and experiment gives new support to the
picture of stripe correlations of holes in La$_{2-x}$ Sr$_x$ Cu O$_4$ for
$0 \leq x \leq 0.018$.

We are deeply indebted to S.A. Kivelson for introducing us to this
problem and for his illuminating comments and suggestions throughout the work.
We thank F. Borsa for discussion of his and other experiments.
We also acknowledge support by NSF Grant PHY 94-07194.

\end{document}